\documentclass[apsrev4-2,prl,twocolumn,footinbib,superscriptaddress,floatfix,bibliography]{revtex4-1}

\usepackage{times}
\usepackage{float}
\usepackage{multirow}
\usepackage[dvipsnames]{xcolor}
\usepackage{amsmath}
\usepackage{amsthm}
\usepackage{amssymb}
\usepackage{amsbsy}
\usepackage{enumitem}
\usepackage{eufrak}
\usepackage{wasysym}
\usepackage[english]{babel}
\usepackage[T1]{fontenc}
\usepackage[utf8]{inputenc} 
\usepackage{graphicx}
\usepackage[colorlinks,bookmarks=false,citecolor=blue,linkcolor=red,urlcolor=blue]{hyperref}
\usepackage[normalem]{ulem}
\usepackage{pstricks}
\usepackage{rotating}			       
\usepackage{tabularx,hhline}	
\usepackage[caption=false]{subfig}		
\newcolumntype{P}[1]{>{\centering\arraybackslash}p{#1}}

\usepackage{pdfpages}

\makeatletter
\AtBeginDocument{\let\LS@rot\@undefined}
\makeatother

\begin{document}


\title{
Spin nematics meet spin liquids: 
Exotic quantum phases in the spin-$1$ bilinear-biquadratic 
model with Kitaev interactions
}


\author{Rico Pohle}
\affiliation{Department of Applied Physics, University of Tokyo, Hongo,  Bunkyo-ku,
Tokyo, 113-8656, Japan}
\author{Nic Shannon}
\affiliation{Theory of Quantum Matter Unit, Okinawa Institute of Science and 
Technology Graduate University, Onna-son, Okinawa 904-0412, Japan} 
\author{Yukitoshi Motome}
\affiliation{Department of Applied Physics, University of Tokyo, Hongo,  Bunkyo-ku,
Tokyo, 113-8656, Japan}

\date{\today}

\begin{abstract}

Spin liquid crystals are magnetic analogs of liquid crystals, possessing properties 
of both liquids and solids, a typical example of which are spin nematics.
%
Spin nematics share many features with spin liquids, and the interplay 
between them is a promising, but little explored, route to uncovering new 
phases of matter.
Here, we address this question  in the context of a spin-$1$ 
magnet on the honeycomb lattice, by considering a model with both 
biquadratic interactions,  favoring spin-nematic states, and 
Kitaev-like interactions, supporting spin liquids.
Accompanying these, 
where dipole and quadrupole moments compete, 
we find a plethora of exotic phases, including 
multiple-$q$ states with nonzero scalar
spin chirality; 
a quasi-one-dimensional coplanar phase; 
a twisted conical phase; 
and a noncoplanar order state which gives way to a chiral spin liquid
at finite temperature.
The implication of these results for experiment is discussed.

\end{abstract}

\maketitle

{\it Introduction---}
%
%
Liquid crystals
form a state of matter 
which contains properties of both liquids and solids~\cite{Kelker1973, Kelker1988, deGennes1993}.
Well-known for their fascinating ``Schlieren'' textures, the nematic phase 
of an organic liquid crystal is made of rod-shaped molecules which align 
themselves along a particular axis, while still flowing like a liquid with the absence 
of crystalline positional order~\cite{Mermin1966, Alexander2012}.
Their magnetic analogs, called spin liquid crystals, 
are spin states in magnets which show similar phenomena.
A typical example is the spin nematic state 
where localized magnetic quadrupole moments act like rod-shaped molecules and form 
orientational order without any long-range magnetic dipole order~
\cite{Blume1969, Matveev1973, Matveev1973, Papanicolaou1988,Tsunetsugu2006, Lauchli2006, Penc2011}.
Such nematic states have attracted much attention as ``hidden magnetic orders'' since they are 
hard to detect with conventional scattering 
experiments~\cite{Andreev1984, Smerald2015, Smerald2016, Kohama2019}.

%
Spin liquids represent another family of spin states which are hard to detect, due to the 
absence of conventional long-range magnetic 
order~\cite{Intro.Frustr.Mag2011, Diep2013, SpinIce2021, Wen2019, Broholm2020}.
Strong competition between magnetic interactions frustrates the system to form a liquid-like 
state made of spins on the lattice. 
Such states are of particular interest, since they are accompanied by emergent gauge fields, 
topological order, and fractionalized excitations, defying a conventional description  
within the Landau paradigm of 
magnetism~\cite{Wen2002, Balents2010, Savary2017, Zhou2017}. 

%
The central question we address in this study is 
``What happens when a spin nematic and a spin liquid meet?''.
While both states stay ``hidden'' for conventional magnetic probes, their 
fundamental properties manifest in different ways:
The spin nematic breaks spin rotation symmetry and shows collective magnetic excitations 
associated with quadrupolar order, whereas the
spin liquid shows no conventional symmetry breaking, with 
magnetic excitations forming a continuum associated with fractional quasiparticles. 
The interplay between spin-nematic and spin-liquid phases offers 
a promising route to uncovering other unconventional phases, and phase transitions, 
in both classical \cite{Chalker1992,Shannon2010,Taillefumier2017}
and quantum spin models  \cite{Shindou2009,Grover2011,Benton2018, Hu2019, Yang2021}.
However, to date this interesting problem remains largely unexplored, because 
of the technical difficulty of solving models which support such ``hidden'' phases.

%
In this Letter, we explore what happens when spin liquids meet spin nematics 
in a spin-1 magnet on the honeycomb lattice.
We consider a bilinear-biquadratic (BBQ) model, extended through anisotropic 
Kitaev-type interactons.
The biquadratic interactions stabilize spin nematic phases with dominant 
quadrupole character~\cite{Zhao2012}.
Meanwhile, the Kitaev interactions stabilize the so-called Kitaev spin liquid~\cite{Kitaev2006}, 
which has been extensively studied for an effective 
spin-orbital entangled moment $S =1/2$~\cite{Jackeli2009, Janssen2019, Takagi2019, Motome2020, Trebst2022},
and its recent extension to higher $S$ in theory and 
experiment~\cite{Baskaran2008, Stavropoulos2019, Scheie2019, Dong2020, Fukui2022, Jin2022, Rayyan2022, Taddei2022}.
We reveal that the competition between 
the positive biquadratic interaction and the Kitaev interaction promotes 
unconventional phases by mixing dipoles and quadrupoles to alleviate 
strong frustration.


%
\begin{figure*}[t]
	\centering
  	\includegraphics[width=0.99\textwidth]{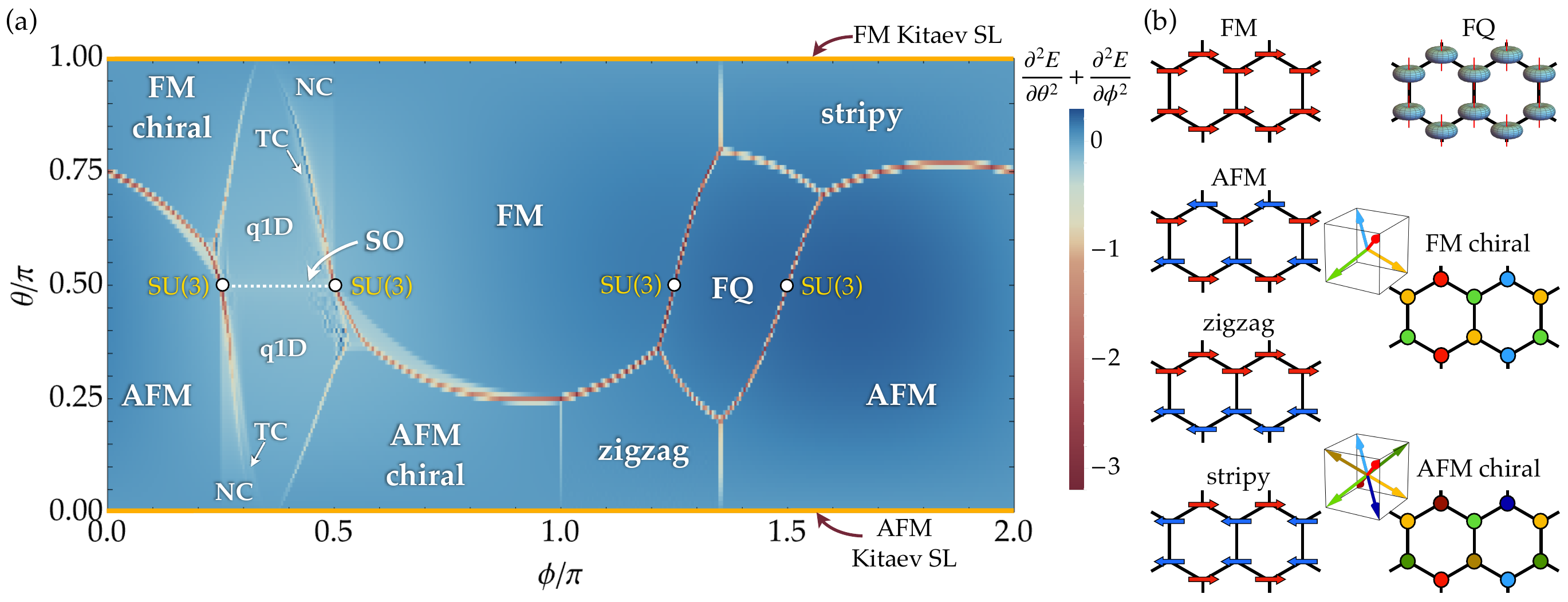}	
	\caption{ 
	Ground states of $\mathcal{H}_{ {\sf BBQ-K} }$ in Eq.~(\ref{eq:H.BBQ-K}), 
	obtained by variational energy minimization for a finite-size cluster with 
	$N=28 \ 800$ spins under periodic boundary conditions. 
	(a) 2D equirectangular projection of the full phase diagram as a function of 
	$\theta$ and 
	$\phi$.
	We plot the second derivative of the internal 
	energy by the contour color to make the phase boundaries visible.
	(b) Real-space configurations of the dominant ordered phases in (a).
	For negative $J_2$ $(1.0 < \phi / \pi < 2.0)$ the model stabilizes 
	magnetic dipolar orders, such as ferromagnetic (FM), antiferromagnetic (AFM), 
	zigzag, and stripy orders, in addition to spin-nematic ferroquadrupolar (FQ) order.
	Meanwhile, for positive $J_2$ $(0.0 < \phi / \pi < 1.0)$ the competition 
	between chiral magnetic order, the Kitaev spin liquid (SL), and 
	quadrupolar semi order (SO) gives rise to unconventional phases, such 
	as the twisted conical (TC), quasi-one-dimensional (q1D) coplanar, 
	and noncoplanar (NC) ordered phases. 
	Details of those unconventional phases will be shown in 
	Fig.~\ref{fig:PhaseDiagram.zoom}.
	}
	\label{fig:PhaseDiagram}
\end{figure*}
%

{\it Ground-state phase diagram.}
%
%
We solve the BBQ model under influence of Kitaev interactions for $S=1$
magnetic moments on the honeycomb lattice:
%
\begin{equation}
	\begin{aligned}
		\mathcal{H}_{ {\sf BBQ-K} } 
			=  \sum_{\langle i,j \rangle} &\left[ \ J_1 \  {\bf S}_i \cdot{\bf S}_j \ 
				+ \ J_2 \ ( {\bf S}_i \cdot {\bf S}_j )^2 \ \right] 	\\
				+ &K  \sum_{\alpha = x,y,z} \  \sum_{ \langle ij \rangle_{\alpha}}  
				S_i^{\alpha} S_j^{\alpha} 	\, ,
		\label{eq:H.BBQ-K} 	
	\end{aligned}
\end{equation}
%
where $J_1$, $J_2$, and $K$ respectively account for the Heisenberg (bilinear), biquadratic, and
Kitaev interaction strengths on nearest-neighbor bonds. 
The index $\alpha$ selects the spin- and bond-anisotropic Kitaev interactions on the 
honeycomb lattice~\cite{Kitaev2006}. 
We parametrize the model by normalizing the total interaction strength as 
%
\begin{equation}
	\left(J_1, J_2, K \right)  = \left( 
		\sin{\theta} \cos{\phi},  
		\sin{\theta} \sin{\phi},
		\cos{\theta} \right) 	\, .
\end{equation}	
%
In this form, $\mathcal{H}_{ {\sf BBQ-K} }$ recovers the well-known limits of 
the BBQ model on the equator, $\theta / \pi= 0.5$ ($K = 0$);
the AFM Kitaev model at the north pole, $\theta / \pi= 0$ ($K = 1$, $J_1 = J_2 = 0$);
the FM Kitaev model at the south pole, $\theta / \pi= 1$ ($K = -1$, $J_1 = J_2 = 0$);
and the Kitaev-Heisenberg model at $\phi / \pi= 0$ or $1$ ($J_2 = 0$).

%
To study the ground-state and finite-temperature properties of 
$\mathcal{H}_{ {\sf BBQ-K} }$ in Eq.~(\ref{eq:H.BBQ-K}),
we utilize a recently developed U(3) formalism~\cite{Remund2022}. 
By embedding the underlying $\mathfrak{su}(3)$ algebra of $S=1$ 
moments  
into the larger $\mathfrak{u}(3)$ 
algebra with an additional spin-length constraint, we are able 
to treat quantum aspects of the problem exactly at the level of a single site, namely, 
simultaneously access dipole and quadrupole fluctuations, despite the drawback 
of loosing quantum entanglement across the lattice
(see technical details in the Supplemental Material~\cite{SM}).
This approach is equivalent to the formalism of SU(3) coherent 
states~\cite{Gnutzmann1998}, which can be generalized to any arbitrary Lie Group 
SU(N)~\cite{Perelomov1972, Zhang2021, Dahlbom2022a, Dahlbom2022b}.

%
In Fig.~\ref{fig:PhaseDiagram}(a) we show the 
ground-state phase diagram of $\mathcal{H}_{ {\sf BBQ-K} }$, obtained from large-scale
variational energy minimization, using the machine learning library JAX \cite{Jax2018, Optax2020} for a 
sufficiently large cluster with $N_{\rm s} = 2L^2= 28,800$ spins ($L=120$)
under periodic boundary conditions.
To visualize phase boundaries, we plot $\partial^2 E/\partial \theta^2 + \partial^2 E/\partial \phi^2$, 
the sum of the second derivatives of the internal energy $E$ with respect to $\theta$ and $\phi$.

Where biquadratic interactions $J_2$ are negative $(1.0 < \phi / \pi < 2.0)$, we recover a combination of 
phases previously reported in the $S=1$ Kitaev-Heisenberg model~\cite{Stavropoulos2019, Dong2020}
and  the BBQ model~\cite{Zhao2012}.
These comprise dipolar order, in the form of ferromagnetic (FM), antiferromagnetic (AFM), 
zigzag, and stripy phases~\cite{Stavropoulos2019, Dong2020}, together with a 
ferroquadrupolar (FQ) spin nematic, stabilized by biquadratic interactions~\cite{Zhao2012}. 
Corresponding real-space spin configurations are shown in Fig.~\ref{fig:PhaseDiagram}(b), 
with dipole moments depicted as arrows, and quadrupole moments as 
``doughnut-shaped'' spin-probability distributions  \cite{Lauchli2006, Tsunetsugu2006}.
%

%
\begin{figure*}[t]
	\centering
  	\includegraphics[width=0.99\textwidth]{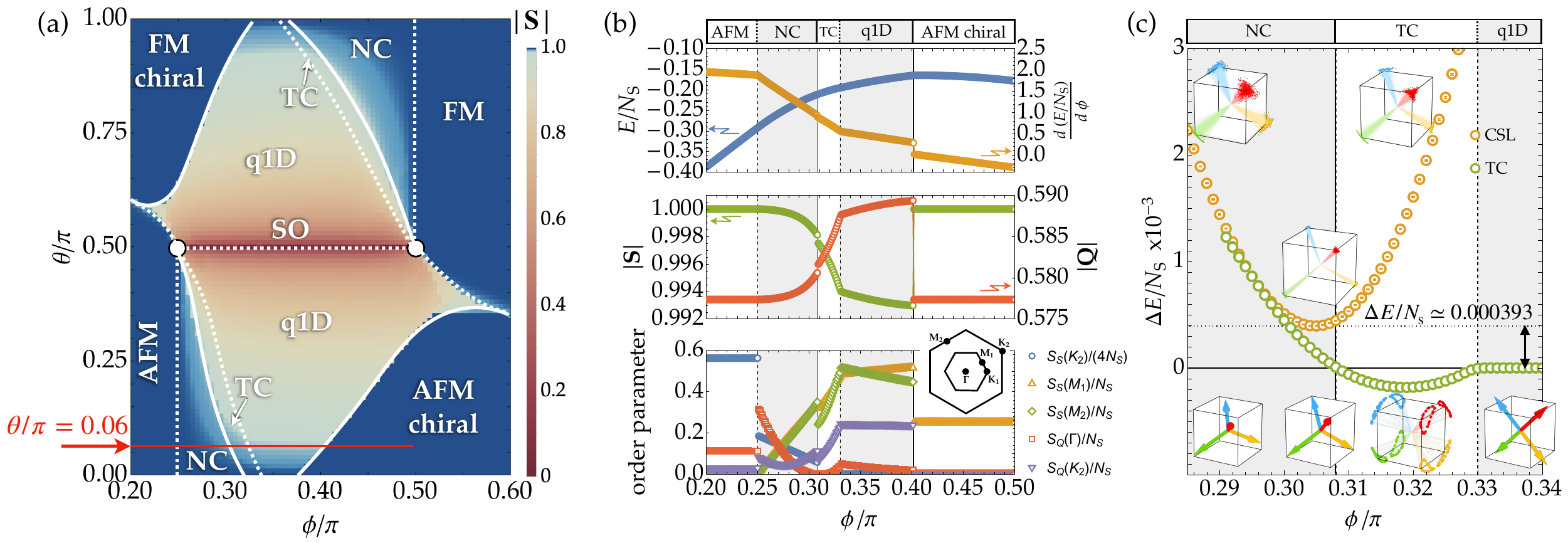}	
	\caption{ 
	Competing region between dipoles and quadrupoles.
	(a) Averaged spin 
	 norm $|{\bf S}|$ plotted in the range of $0.2 < \phi / \pi < 0.6$. 
	The solid and dashed lines represent first- and second-order 
	phase boundaries, respectively.
	(b) Normalized energy and its derivative, $E/N_{\rm s}$ and $\partial (E/N_{\rm s}) / \partial \phi$, 
	respectively, which visualize the boundaries between the antiferromagnet (AFM), 
	noncoplanar order (NC), twisted conical (TC), quasi-one-dimensional (q1D) coplanar, 
	and AFM chiral states (top); 
	spin and quadrupole  
	norms, $|{\bf S}|$ and $|{\bf Q}|$, respectively (middle),
	and the structure factors for dipole and quadrupole moments, 
	$S_{\rm{S}}({\bf q})$ and $S_{\rm{Q}}({\bf q})$, respectively, 
	at high-symmetry momenta (bottom) at $\theta/\pi = 0.06$ [red line in (a)].
	(c) Energy measured from the NC state at $\theta/\pi = 0.06$. 
	The lower insets show spin configurations 
	in each ordered phase, while the upper ones represent metastable states
	of a chiral spin liquid (CSL).
	}
	\label{fig:PhaseDiagram.zoom}
\end{figure*}
%

In contrast, where the biquadratic interaction $J_2$ is positive $(0.0 < \phi / \pi < 1.0)$, 
we find a number of unconventional phases which emerge from the competition between 
frustrated interactions.
%
Adjacent to the zigzag and stripy phases, we find FM and AFM chiral phases, which are
noncoplanar dipolar-ordered phases with net scalar spin chirality, 
$|\kappa | = 8/ (3 \sqrt{3}) $ and $| \kappa | = 16/(3 \sqrt{3})$ in the regions 
for $\theta/\pi>0.5$ and $\theta\pi<0.5$, respectively
(see definition of $\kappa$ in the Supplemental Material~\cite{SM}).
These chiral phases are triple-$q$ states with eight-sublattice order, also
known as ``tetrahedral'' and ``cubic'' states~\cite{Messio2011}, 
in which spins point to the corners of a unit cube [see Fig.~\ref{fig:PhaseDiagram}(b)].
\footnote{
We notice several complicated phases around the two SU(3) points 
at $(\theta/\pi, \phi/\pi) = (0.5, 0.25)$ and $(0.5, 0.5)$.  
These will be discussed elsewhere.}.
We note that the stripy and FM chiral (zigzag and AFM chiral) states are energetically degenerate along the 
vertical line at $\phi/\pi=0$ ($\phi/\pi = 1$) for $\theta/\pi \geq 0.75$ ($\theta/\pi \leq 0.25$).
This degeneracy is lifted by thermal fluctuations, which select the stripy (zigzag) phase, 
consistent with DMRG results for the spin--1 Kitaev--Heisenberg model \cite{Dong2020}.


{\it Competing region between dipoles and quadrupoles.}
Between FM, AFM and chiral ordered phases, dipole and quadrupole 
moments mix in nontrivial ways, and we find a range of exotic phases.
These results are summarised in Fig.~\ref{fig:PhaseDiagram.zoom}(a), 
where the color bar shows the expectation value of the magnitude of the dipole moment 
$|{\bf S}| = \sum_i |{\bf S}_i |/N_{\rm s}$. 
%
%
%
In the
limit of the BBQ model 
($\theta/\pi = 0.5$), 
a purely quadrupolar state with $|{\bf S}| = 0$ is realized,
connecting the two SU(3) points at $\phi/\pi = 0.25$ and $\phi/\pi = 0.5$. 
This state is known to be semi ordered (SO) 
(or semi-disordered)~\cite{Papanicolaou1988, Niesen2017}, 
with $\langle {\bf S}_i \cdot {\bf S}_j \rangle = 0$ for all the bonds to minimize 
$\mathcal{H}_{\sf BBQ}$ with positive $J_1$ and $J_2$. 
We note that this state 
was claimed to be replaced by
a plaquette valence bond crystal 
when quantum entanglement is fully taken into account~\cite{Corboz2013}.
%
%
Introducing the Kitaev interaction $|K|$ ($\theta/\pi \neq 0.5$) immediately induces the 
formation of a nonzero dipole moment $|{\bf S}| \neq 0$. 
Under the orthogonal 
conditions of $\langle {\bf S}_i \cdot {\bf S}_j \rangle = 0$, 
the Kitaev interaction prefers a coplanar state with dipoles aligned within the 
$xy$, $yz$, or $zx$ plane, in which the Kitaev energy is minimized on two 
bonds while leaving the third one to be zero.
Hence, dominant correlations prevail  along isolated chains and 
stabilize a coplanar, quasi-one-dimensional (q1D) state in an extended region away from the BBQ limit
[see Fig.~\ref{fig:PhaseDiagram.zoom}(a)].
The appearance of small quadrupolar correlations will induce a weak interchain coupling to 
form two-dimensional order. 
%

Sandwiched between the q1D coplanar phase and the AFM/FM ordered phase, 
$\mathcal{H}_{\sf BBQ-K }$ minimizes its energy by forming 
spin textures with noncoplanar (NC) orientation. 
In Fig.~\ref{fig:PhaseDiagram.zoom}(b) we show physical observables along a cut with $\theta/\pi = 0.06$ 
[red line in Fig.~\ref{fig:PhaseDiagram.zoom}(a)], visiting AFM, NC, twisted conical (TC), 
q1D coplanar, and AFM chiral phases. 
Phase boundaries are easily distinguished from the first derivative of energy $d (E/N_{\rm s}) / d \phi$ 
[top panel of Fig.~\ref{fig:PhaseDiagram.zoom}(b)].
Each phase has been identified from the spin and quadrupole norms, 
$|{\bf S}|$ and $|{\bf Q}|$ shown in the middle panel of Fig.~\ref{fig:PhaseDiagram.zoom}(b), and
their order parameters, measured from the structure 
factors $S_{\rm{S}}({\bf q})$ for dipoles and $S_{\rm{Q}}({\bf q})$ for quadrupoles 
at high symmetry momenta, $\Gamma$, $K_1$, $K_2$, $M_1$, and $M_2$, shown in 
the bottom panel of Fig.~\ref{fig:PhaseDiagram.zoom}(b)
(see definitions in the Supplemental Material~\cite{SM}).

By increasing $\phi/\pi > 0.25$, spins start to cant away from the 
collinear AFM arrangement until they point almost along the corners 
of a unit cube for $\phi/\pi \approx 0.305$, as shown in the lower inset of Fig.~\ref{fig:PhaseDiagram.zoom}(c). 
For $\phi/\pi \gtrsim 0.308$, the ground state becomes a four-sublattice 
conical state, where spins process around two corners of a unit cube. 
This precession, however, is not circular, but rather shows a twist with 
a node between those two corners, which led us to call this state TC.
Further increase of $\phi$ makes the cones shrink 
and smoothly merge with the q1D coplanar state at $\phi/\pi \gtrsim 0.330$.
Near the 
transition between the NC and TC phases, we observe 
a metastable noncoplanar state in the energy minimization, whose 
energy minimum is just 
$\Delta E / N_{\rm s} \simeq 0.000393$ 
above the NC ground state.
Interestingly, we obtain many different noncoplanar spin states with almost the same energy, 
which, in fact, form 
an extensively degenerate chiral spin liquid (CSL).


%
\begin{figure}[t]
	\centering
  	\includegraphics[width=0.49\textwidth]{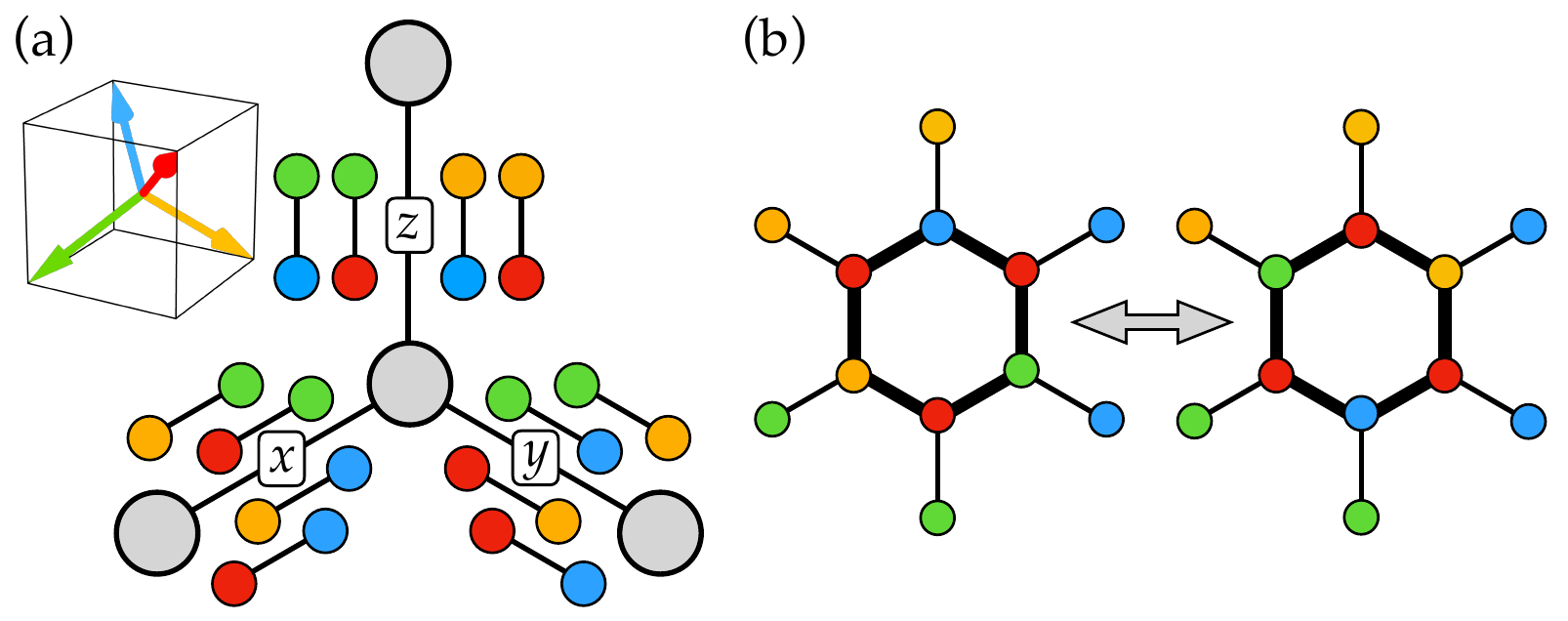}	
	\caption{ 
	Macroscopic degeneracy in the eight-color chiral spin 
	liquid (CSL).
	(a) Spin and bond constraints allow for only four color pairs on 
	individual Kitaev bonds.
	Inset assigns colors to spins perfectly pointing 
	to the corners of the unit cube.
	(b) While respecting the constraints in (a) the CSL manifold allows for 
	pairs of spin states per hexagon, which are related by a six-spin cluster update.
	All allowed hexagon configurations are shown in the Supplemental Material~\cite{SM}.
	}
	\label{fig:CSL.8color.model}
\end{figure}
%

{\it Eight-color chiral spin liquid.}
The extensive degeneracy of the CSL is transparently understood by two simplifications. 
One  is to assume $|{\bf S}| = 1$, the other is to allow for only eight discrete spin states 
which exactly point to the corners of a unit cube.
Such simplifications are justified since the metastable states are dominantly made of dipoles 
and their spins gather around the corners, as seen in the upper insets 
of Fig.~\ref{fig:PhaseDiagram.zoom}(c).
Under these simplifications, positive
$J_1$ and $J_2$ 
select only four out of eight discrete spin states,
which we identify by color, as indicated in the inset of 
Fig.~\ref{fig:CSL.8color.model}(a) (or equivalently the other
four spins with opposite directions).
Then, the Kitaev interactions enforce bond constraints, 
which allow for only four combinations 
of color pairs per bond, where their explicit combination differs between the three 
Kitaev bonds, as illustrated in Fig.~\ref{fig:CSL.8color.model}(a).
These local bond constraints are not strong enough to enforce long-range order, instead 
form a manifold with macroscopic degeneracy, always allowing for two spin configurations 
per hexagon [see Fig.~\ref{fig:CSL.8color.model}(b)].
The number of states respecting the constraints in  Fig.~\ref{fig:CSL.8color.model}(a) 
per hexagon is $64$ (see the Supplemental Material~\cite{SM}), leading 
to a residual entropy of 
$S/N_{\rm s} = \frac{1}{2} \log{2}$, which is  $1/6$ of  
the total entropy of the 
system, $S/N_{\rm s} = \log 8$. 
We find that the degenerate manifold gives rise to a nonzero scalar spin chirality; 
$|\kappa| = 1/\sqrt{3}$, 
hence justifying its 
name as the eight-color CSL. 
Since the CSL is a metastable state [see Fig.~\ref{fig:PhaseDiagram.zoom}(c)], 
we expect an entropy-driven phase transition from the NC state to the eight-color CSL
by raising the temperature.
This motivates us to perform finite-temperature Monte Carlo simulations of 
$\mathcal{H}_{ {\sf BBQ-K} }$ within the U(3) formalism, henceforth
called ``u3MC''~\cite{Remund2022}
(see technical details in the Supplemental Material~\cite{SM}).
%


%
\begin{figure}[t]
	\centering
  	\includegraphics[trim={0cm 0 0cm 0}, width=0.49\textwidth]{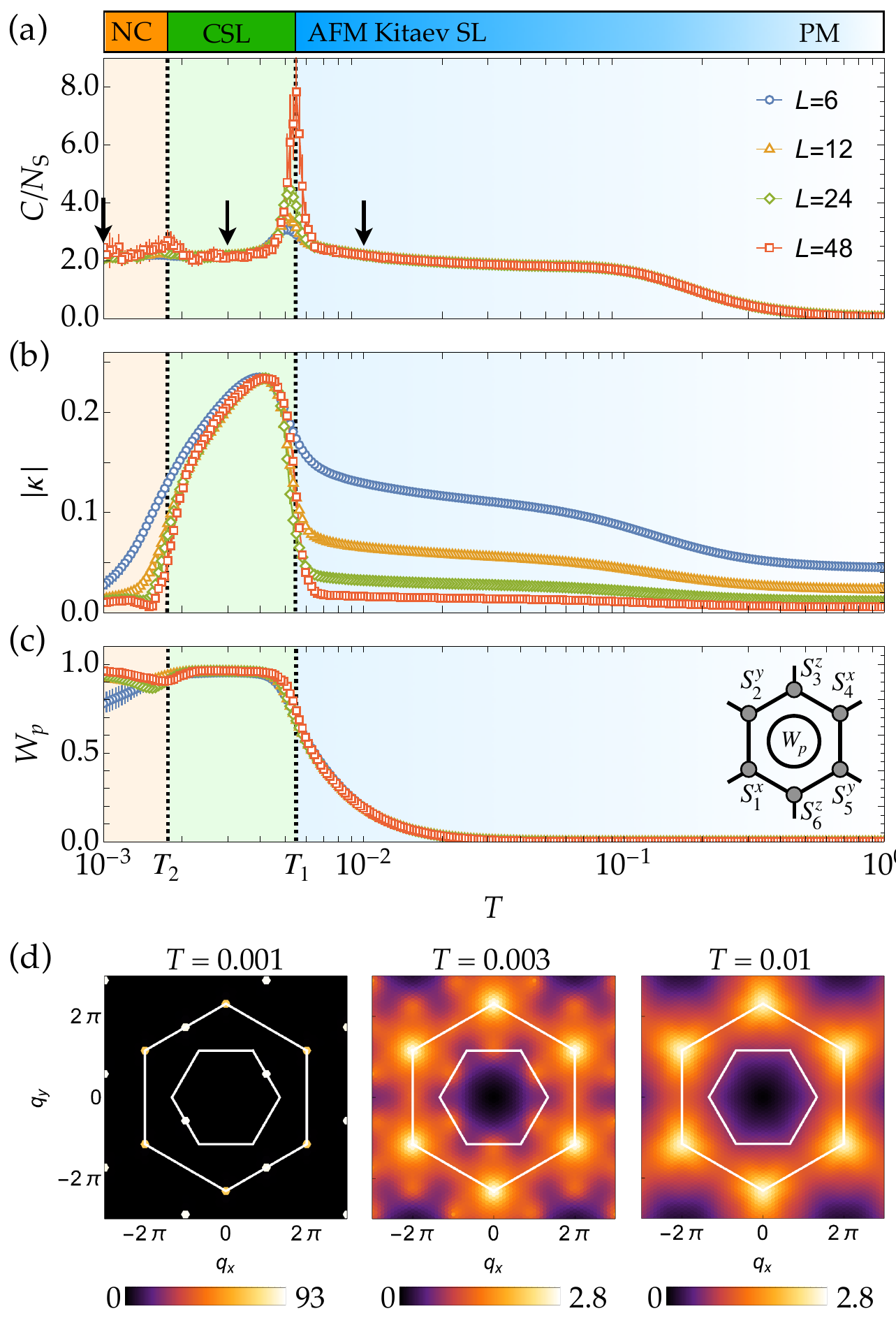}	
	\caption{ 
	Finite-temperature Monte Carlo results at $\theta/\pi = 0.06$ and $\phi / \pi = 0.305$ 
	where the ground state is NC ordered. 
	Shown are $T$ dependences of 
	(a) the specific heat per site, 
	(b) the absolute value of the scalar spin chirality per site, $|\kappa|$ , and
	(c) the $S=1$ semiclassical analog of the $Z_2$-flux operator.
	The data are taken for $L = 6$, $12$, $24$, and $48$ 
	($N_{\rm s} = 72$, $288$, $1152$, and $4608$). 
	(d) Spin structure factor $S_{\rm{S}}({\bf q})$ at temperatures 
	indicated with black arrows in (a), for $L = 12$ at $T=0.001$ and 
	$L = 24$ at $T=0.003$ and $0.01$. 
	}
	\label{fig:Thermo.CSL}
\end{figure}
%

%
In Fig.~\ref{fig:Thermo.CSL} we show 
u3MC results at
$\theta / \pi = 0.06 $ and $\phi / \pi = 0.305$, where the energy difference $\Delta E$ 
to the NC ground state shows a minimum [see Fig.~\ref{fig:PhaseDiagram.zoom}(c)].
The specific heat in Fig.~\ref{fig:Thermo.CSL}(a) shows two singularities at $T_1 =  0.0055(2)$ and 
$T_2 = 0.0017(2)$.
Below $T_1$ we obtain a nonzero scalar spin chirality,
while we do not see any Bragg peaks in the structure factors for both dipole and quadrupole 
sectors; see the dipole one in the middle panel of Fig.~\ref{fig:Thermo.CSL}(d)
and the quadrupole one in the Supplemental Material~\cite{SM}.
Hence, we associate $T_1$ with a discrete $Z_2$ chiral symmetry breaking 
from the high-temperature cooperative paramagnetic state, which retains a 
classical AFM Kitaev spin liquid feature 
as shown in the right panel of Fig.~\ref{fig:Thermo.CSL}(d) \cite{Samarakoon2017}
(see the Supplemental Material~\cite{SM}), into the CSL.
The bond constraints, explained in Fig.~\ref{fig:CSL.8color.model}, lead us to compute
the $S=1$ semiclassical analog of the $Z_2$-flux operator~\cite{Kitaev2006}:
$W_p =  \prod_{j \in p} \sqrt{3} S_j^{\mu}$ on hexagons $p$. 
As plotted in Fig.~\ref{fig:Thermo.CSL}(c), $W_p$ rapidly increases below $T\sim 0.01$ and 
becomes nearly one in the CSL below $T_1$, suggesting that the CSL accompanies ``flux order'' 
like in the Kitaev SL.
With further decrease of temperature, below $T_2$, we find characteristic Bragg peaks corresponding to NC order as shown 
in the left panel of Fig.~\ref{fig:Thermo.CSL}(d), and hence, associate $T_2$ with a symmetry 
breaking into the NC ordered phase. 
Although the estimate of $T_2$ is not fully reliable because of the slowing down 
of u3MC simulations in the low temperature regime, it compares well with the 
free energy $\frac{\Delta E}{S} \approx  0.001 \approx T_2$, 
given $\Delta E/N_{\rm s} \simeq 0.000393$ and the entropy of the CSL manifold
$S/N_{\rm s} = \frac12 \log 2 \simeq 0.347$. 
Thus, we identify the CSL to be 
driven by the entropy associated with the macroscopic degeneracy discussed in 
Fig.~\ref{fig:CSL.8color.model}.
We further confirmed that the intermediate-temperature CSL 
prevails in an extended region around 
$\phi/\pi \sim 0.3$
and 
$0 < \theta/\pi \lesssim 0.2$ (see the Supplemental Material~\cite{SM}).

{\it Conclusions and discussions.}
To answer the question “What happens when a spin nematic and a spin liquid meet?”,
we have theoretically investigated the spin $S = 1$ bilinear-biquadratic model 
on the honeycomb lattice with Kitaev-type anisotropic interactions.
In the pure Kitaev limit we recovered the semiclassical analog of the 
$S=1$ Kitaev spin liquid, which was  
suggested to emerge in 
honeycomb materials with Ni$^{2+}$ ions \cite{Stavropoulos2019}.
The limit of strong negative biquadratic interactions revealed a ferroquadrupolar 
spin nematic state, which can be expected in materials with spin-phonon 
coupling \cite{Tchernyshyov2011, Stoudenmire2009, Valentine2020}.
By tuning model parameters to positive biquadratic interactions,
which are expected in materials with orbital 
degeneracy~\cite{Yoshimori1981, Hoffmann2020, Soni2022},
we observed the formation of triple-$q$ ordered phases with 
net scalar spin chirality. 
Note that a similar state was found for the Kitaev-Heisenberg 
model in a magnetic field \cite{Janssen2017} and the honeycomb 
material Na$_2$Co$_2$TeO$_6$ \cite{Yao2022, Chen2021, Kruger2022}.
Deep in the frustrated region, spin dipole and quadrupole components mix and promote
quasi-one-dimensional coplanar ordered phases, unconventional noncoplanar and conical phases,
and a finite-temperature eight-color chiral spin liquid 
phase retaining the Kitaev spin liquid feature of flux order.
Such exotic phases 
survive for large Kitaev anisotropy offering an opportunity for them to emerge 
as perturbation in higher-$S$ Kitaev magnets
\cite{Stavropoulos2019, Fukui2022, Rayyan2022}.

The results in this Letter were derived within a semiclassical approximation which does 
not take into account entanglement between spins. 
A drawback of this approach is that Kitaev-like spin-liquid behavior is found only exactly 
at the north and south poles, while in a fully quantum treatment, it is expected to be an extended phase
~\footnote{We have confirmed that our method reproduces the 
short-range spin correlations expected for the $S=1$ Kitaev spin liquid \cite{Baskaran2008}
(see also Supplemental Material~\cite{SM}).}.  
None the less, comparison with DMRG results for the spin-$1$ Kitaev-Heisenberg model \cite{Dong2020} reveals 
an encouraging level of agreement, with the same phases identified, and only small changes in the numerical values 
of phase boundaries.


While the $S = 1/2$ Kitaev model offers a bond-nematic state near the Kitaev spin liquid 
\cite{Nasu2017}, the physics in $S = 1$ Kitaev magnets appears to be much 
richer.
The enlarged local Hilbert space allows for on-site multipole fluctuations, which work 
as an additional degree of freedom to minimize local bond energies by forming 
unconventional quantum states of matter.
Similar exotic examples have recently been shown to exist in
${\sf CP}^2$ skyrmion crystals \cite{Amari2022, Zhang2022}, 
on triangular magnets \cite{Bai2021, Seifert2022}, 
and in spin-orbital Mott insulators \cite{Iwazaki2023}.
The possibilities to find novel phases in multipolar magnets are vast and 
the recent development of suitable numerical methods
\cite{Remund2022, Zhang2021, Dahlbom2022a, Dahlbom2022b} allows us to 
access and study them in a controlled way. 
We hope our findings will stimulate further theoretical and experimental 
exploration of multipolar magnets and their exotic quantum phases formed from 
the interplay between spin nematics and spin liquids.


{\it Acknowledgments.}
The authors are pleased to acknowledge helpful conversations with 
Yutaka Akagi,
Yuki Amari,
Koji Inui,
Lukas Janssen, 
Yasuyuki Kato,
and Kotaro Shimizu.
This work was supported by Japan Society for the Promotion of Science 
(JSPS) KAKENHI Grants No. JP19H05822 and No. JP19H05825, and the 
Theory of Quantum Matter Unit, OIST.
Numerical calculations we carried out using HPC facilities provided by 
the Supercomputer Center of the Institute for Solid State Physics, 
the University of Tokyo.

\bibliography{Bibliography}

\clearpage
\widetext


\section*{Supplementary material}

Rico Pohle,[1] Nic Shannon,[2] and Yukitoshi Motome[1]

[1]*Department of Applied Physics, University of Tokyo, Hongo, Bunkyo-ku, Tokyo, 113-8656, Japan*
[2]*Theory of Quantum Matter Unit, Okinawa Institute of Science and Technology Graduate University, Onna-son, Okinawa 904-0412, Japan*
(Dated: March 23, 2023)

## I. DETAILS OF THE METHOD

### A. Description of spin-1 magnets within the U(3) formalism

Magnets made of spin-1 moments contain magnetic dipole and quadrupole components. To correctly account for both of these components, we adopt the recently developed U(3) formalism [1], which embeds the $\mathfrak{su}(3)$ algebra of a $S=1$ spin into the larger $\mathfrak{u}(3)$ algebra with an additional spin-length constraint. This approach is suitable for both analytical and numerical investigations of spin-1 magnets, and allows us to access, next to thermodynamic properties, also their excitations at finite temperatures.

We express a spin-1 moment at site $i$ by the $3\times 3$ Hermitian matrix, $\mathcal{A}_{i\beta}^{\alpha}$, which we call henceforth $\mathcal{A}$ matrix. This $\mathcal{A}$ matrix incorporates informations of dipole and quadrupole moments on site $i$, which can be obtained respectively via

$$S_i^\alpha = -i\epsilon^{\alpha\,\gamma}_{\ \beta}\mathcal{A}_{i\gamma}^{\beta}, \tag{S1}$$

$$Q_i^{\alpha\beta} = -\mathcal{A}_{i\beta}^{\alpha} - \mathcal{A}_{i\alpha}^{\beta} + \frac{2}{3}\delta^{\alpha\beta}\mathcal{A}_{i\gamma}^{\gamma}. \tag{S2}$$

$\epsilon^{\alpha\,\gamma}_{\ \beta}$ is the Levi-Civita symbol and $\delta^{\alpha\beta}$ is the Kronecker delta. We adopt the Einstein convention of summing over repeated indices, and express spin quadrupoles in their symmetric and traceless rank-2 tensor form. By using Eqs. (S1) and (S2) we rewrite Eq. (1) of the main text in terms of $\mathcal{A}$ matrices and obtain a Hamiltonian which is bilinear in $\mathcal{A}_{i\beta}^{\alpha}$

$$\mathcal{H}_{\text{BBQ-K}} = \sum_{\langle ij\rangle}\left[J_1\mathcal{A}_{i\beta}^{\alpha}\mathcal{A}_{j\alpha}^{\beta} + (J_2-J_1)\mathcal{A}_{i\beta}^{\alpha}\mathcal{A}_{j\beta}^{\alpha} + J_2\mathcal{A}_{i\alpha}^{\alpha}\mathcal{A}_{j\beta}^{\beta}\right] - K\sum_{\langle ij\rangle_\alpha}\epsilon^{\alpha\,\gamma}_{\ \beta}\epsilon^{\alpha\,\eta}_{\ \delta}\mathcal{A}_{i\gamma}^{\beta}\mathcal{A}_{j\eta}^{\delta}. \tag{S3}$$

For convenience, we extract dipole and quadrupole components from the $\mathcal{A}$ matrices directly via

$$\begin{pmatrix}\mathbf{S}^2\\S^x\\S^y\\S^z\\Q^{x^2-y^2}\\Q^{3z^2-r^2}\\Q^{xy}\\Q^{xz}\\Q^{yz}\end{pmatrix} = C\begin{pmatrix}\mathcal{A}_1^1\\\mathcal{A}_2^1\\\mathcal{A}_3^1\\\mathcal{A}_1^2\\\mathcal{A}_2^2\\\mathcal{A}_3^2\\\mathcal{A}_1^3\\\mathcal{A}_2^3\\\mathcal{A}_3^3\end{pmatrix}, \tag{S4}$$

with the spin length constraint $\mathbf{S}^2 = S(S+1) = 2$, and $C$ the $9\times 9$ transformation matrix:

$$C = \begin{pmatrix}2 & 0 & 0 & 0 & 2 & 0 & 0 & 0 & 2\\ 0 & 0 & 0 & 0 & 0 & -i & 0 & i & 0\\ 0 & 0 & i & 0 & 0 & 0 & -i & 0 & 0\\ 0 & -i & 0 & i & 0 & 0 & 0 & 0 & 0\\ -1 & 0 & 0 & 0 & 1 & 0 & 0 & 0 & 0\\ \frac{1}{\sqrt{3}} & 0 & 0 & 0 & \frac{1}{\sqrt{3}} & 0 & 0 & 0 & -\frac{2}{\sqrt{3}}\\ 0 & -1 & 0 & -1 & 0 & 0 & 0 & 0 & 0\\ 0 & 0 & -1 & 0 & 0 & 0 & -1 & 0 & 0\\ 0 & 0 & 0 & 0 & 0 & -1 & 0 & -1 & 0\end{pmatrix}. \tag{S5}$$

To compute the averaged spin-dipole and spin-quadrupole norm we respectively use

$$|\mathbf{S}| = \frac{1}{N_{\text{s}}}\sum_i |\mathbf{S}_i|, \tag{S6}$$

$$|\mathbf{Q}| = \frac{1}{N_{\text{s}}}\sum_i |\mathbf{Q}_i|, \tag{S7}$$

where vector components are expressed as

$$\mathbf{S}_i = \begin{pmatrix}S_i^x\\S_i^y\\S_i^z\end{pmatrix}, \tag{S8}$$

and

$$\mathbf{Q}_i = \begin{pmatrix}Q_i^{x^2-y^2}\\Q_i^{3z^2-r^2}\\Q_i^{xy}\\Q_i^{xz}\\Q_i^{yz}\end{pmatrix} = \begin{pmatrix}\frac{1}{2}(Q_i^{xx}-Q_i^{yy})\\ \frac{1}{\sqrt{3}}(Q_i^{zz}-\frac{1}{2}(Q_i^{xx}+Q_i^{yy}))\\Q_i^{xy}\\Q_i^{xz}\\Q_i^{yz}\end{pmatrix}. \tag{S9}$$

In our numerical treatment we express $\mathcal{A}$ matrices

$$\mathcal{A}_{i\beta}^{\alpha} = (\mathrm{d}_i^{\alpha})^*\,\mathrm{d}_{i\beta}, \tag{S10}$$

in terms of complex vectors $\mathbf{d}_i$

$$\mathbf{d} = \begin{pmatrix} x_1 + i\, x_2 \\ x_3 + i\, x_4 \\ x_5 + i\, x_6 \end{pmatrix}, \quad \text{(S11)}$$

which are also known as directors [1–3]. We implicitly respect the spin-length constraint on the directors

$$\mathbf{d}^*\mathbf{d} = |\mathbf{d}|^2 = 1, \quad \text{(S12)}$$

by parametrizing the components of $\mathbf{d}$ on a five-dimensional sphere

$$\begin{aligned}
x_1 &= \theta_2^{1/4}\, \theta_1^{1/2}\, \sin\phi_1\,, \\
x_2 &= \theta_2^{1/4}\, \theta_1^{1/2}\, \cos\phi_1\,, \\
x_3 &= \theta_2^{1/4}\, (1-\theta_1)^{1/2}\, \sin\phi_2\,, \\
x_4 &= \theta_2^{1/4}\, (1-\theta_1)^{1/2}\, \cos\phi_2\,, \\
x_5 &= (1-\theta_2^{1/2})^{1/2}\, \sin\phi_3\,, \\
x_6 &= (1-\theta_2^{1/2})^{1/2}\, \cos\phi_3\,,
\end{aligned} \quad \text{(S13)}$$

with $0 \leq \theta_1, \theta_2 \leq 1$ and $0 \leq \phi_1, \phi_2, \phi_3 < 2\pi$. To fix the unphysical gauge one can set $\phi_3 = \pi/2$, which will make the $z$ component of $\mathbf{d}$ purely real, providing in total four degrees of freedom as anticipated for a spin-1 moment [1].

### B. Variational energy minimization

To obtain the ground-state phase diagram, as shown in Fig.1 of the main text, we performed large-scale variational energy minimization, using the machine learning library JAX [4]. We took $\theta_1, \theta_2, \phi_1, \phi_2$, and $\phi_3$ as variational parameters, and minimized the energy for $\mathcal{H}_{\mathsf{BBQ-K}}$ in Eq. (S3). The energy minimization has been done with the gradient processing and optimization library "Optax", using the optimizer "Adam" [5]. As the initial conditions, we considered dipole and quadrupole ordered states, disordered chiral spin liquid (CSL) states, and random states. The optimization has been done for 5000-10000 steps per model parameter.

### C. Monte Carlo simulation

To obtain thermodynamic properties we performed semiclassical Monte Carlo (MC) simulations using the U(3) formalism, also known as "u3MC" [1]. Simulations are considered as semiclassical in the sense that the quantum mechanics of each local $S = 1$ moment is treated exactly at the level of a single site, however, nonlocal quantum entanglement beyond a single site is neglected [6–10].

Within the u3MC simulations, the $\mathcal{A}$ matrices are sampled using the Marsaglia construction [11] at every site on the physical lattice of size $N_\mathrm{s}$ by choosing parameters $0 \leq \theta_1, \theta_2 \leq 1$ and $0 \leq \phi_1, \phi_2, \phi_3 < 2\pi$ in Eq. (S13) at random. Following the standard single-spin flip Metropolis algorithm [12] we accept or reject a new $\mathcal{A}$ matrix state at site $i$, after evaluating the local bond energies from Eq. (S3). In the actual calculations, a single MC step consists of $N_\mathrm{s}$ local spin-flip attempts on randomly chosen sites. Simulations were performed in parallel for replicas at different temperatures, using the replica-exchange, initiated by the parallel tempering algorithm [13–15] every 100 MC steps. Results for thermodynamic quantities were averaged over $5 \times 10^5$ statistically independent samples, after initial $5 \times 10^5$ MC steps for slowly heating from the ground state, as obtained by the variational energy minimization, and further $5 \times 10^5$ MC steps for thermalization. Error bars were estimated by averaging over 10 independent simulation runs.

MC simulations in the frustrated parameter region, where the eight-color CSL appears at finite temperature, suffer from severe slowing down, due to the failure of single-spin flip updates. As explained in the main text, the CSL is expected to have an extensively large number of energetically degenerate states, which are connected via simultaneous flips of a group of spins covering at least one hexagon, as exemplified in Fig. 3(b) of the main text. To efficiently sample over such states, we implement a cluster update of six spins by

$$\mathcal{A}_j^{\mathsf{new}} = R_\alpha^{-1} \mathcal{A}_j R_\alpha \quad \text{for } j \in p, \quad \text{(S14)}$$

where $R_\alpha$ is the rotation matrix of angle $\pi$ about the axis $\alpha = x, y, z$. $\alpha$ is chosen site-dependent as the label for the bond of Kitaev interactions pointing outside the hexagon $p$ on site $j$. By comparing energies before and after the cluster update, we accept or reject the update following the Metropolis argument [12]. Within one MC step we implement $N_{\mathsf{hex}} = N_\mathrm{s}/2$ cluster-update attempts for randomly chosen hexagons on the whole lattice.

## II. EIGHT-COLOR CHIRAL SPIN LIQUID

### A. Spin configurations and scalar spin chirality

As stated in the main text, the scalar spin chirality in the eight-color CSL remains nonzero. We define the scalar spin chirality on the honeycomb lattice with

$$\kappa = \frac{2}{N_\mathrm{s}} \sum_p \kappa_p\,, \quad \text{(S15)}$$

over all hexagons $p$ on the lattice. As shown in Fig. S1 we divide a single hexagon into four triangles which contribute to

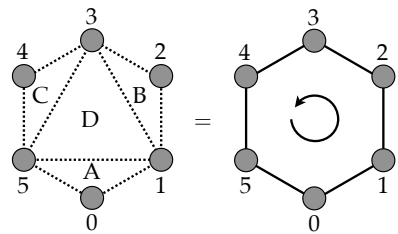

Figure S1. Definition of the scalar spin chirality on an individual hexagon, see Eq. (S16).

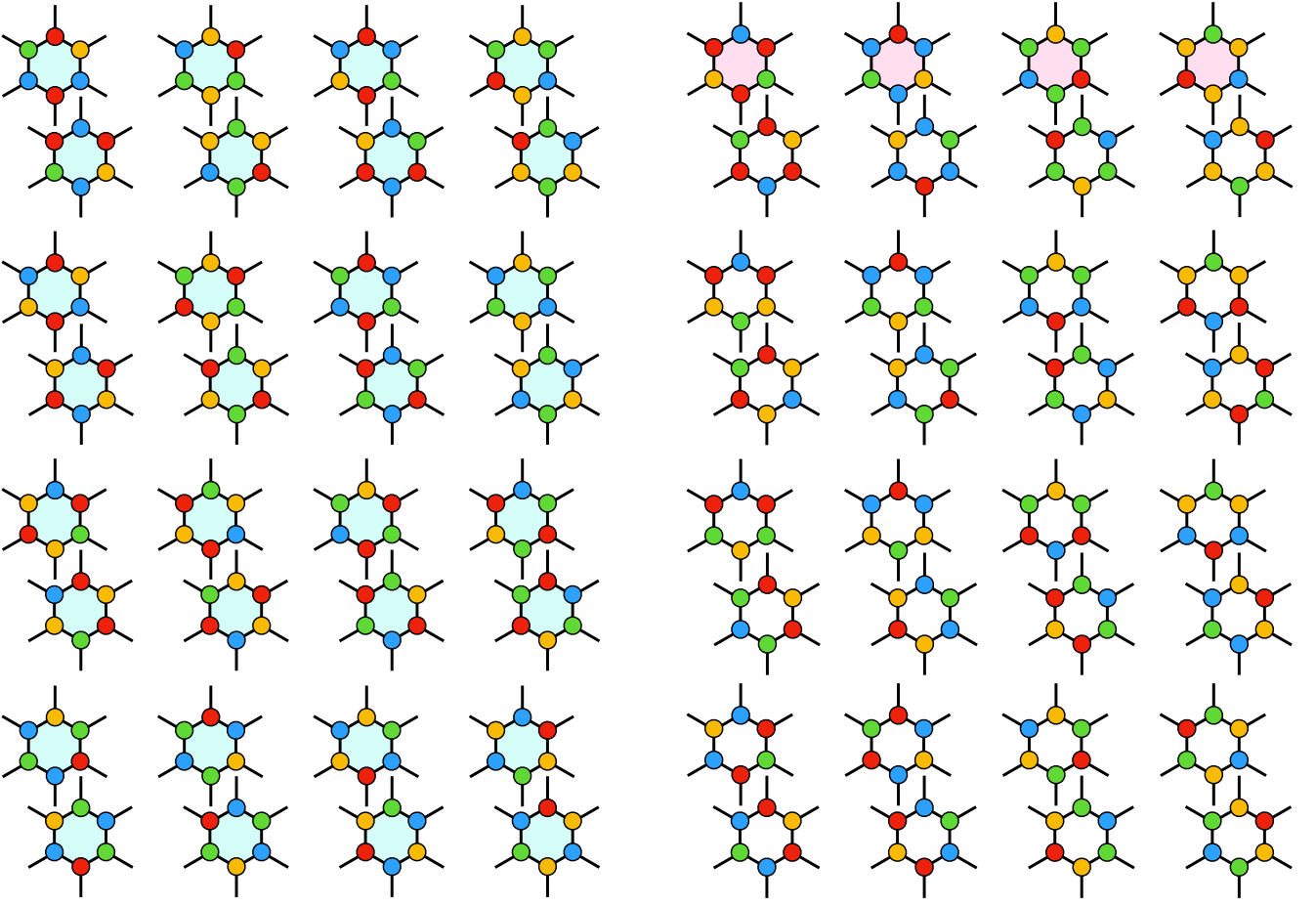

Figure S2. All allowed spin arrangements on a single hexagon in the eight-color chiral spin liquid. The definition for colors are the same as those in Fig. 3 of the main text. The paired configurations are related to each other by a six-site hexagon update, as exemplified in Fig. 3(b) of the main text. Hexagons are shaded in light cyan, pink, and white, indicating the value of scalar spin chirality: $\kappa_p = -8/(3\sqrt{3})$, $16/(3\sqrt{3})$, and 0, respectively.

the total scalar chirality of the full hexagon with

$$\kappa_p = \kappa_p^A + \kappa_p^B + \kappa_p^C + \kappa_p^D, \tag{S16}$$

where

$$\kappa_p^A = \mathbf{S}_{p0} \cdot (\mathbf{S}_{p1} \times \mathbf{S}_{p5}), \tag{S17}$$
$$\kappa_p^B = \mathbf{S}_{p1} \cdot (\mathbf{S}_{p2} \times \mathbf{S}_{p3}), \tag{S18}$$
$$\kappa_p^C = \mathbf{S}_{p3} \cdot (\mathbf{S}_{p4} \times \mathbf{S}_{p5}), \tag{S19}$$
$$\kappa_p^D = \mathbf{S}_{p1} \cdot (\mathbf{S}_{p3} \times \mathbf{S}_{p5}). \tag{S20}$$

There are 64 energetically degenerate spin configurations per hexagon that satisfy the bond constraints of the CSL state. The complete list of allowed states is displayed in Fig. S2; see definition of spin coloring and bond constraints in Fig. 3 of the main text. We show configurations in pairs, which are related by a six-site hexagon update with Eq. (S14), as exemplified in Fig. 3(b) of the main text. This degenerate manifold has three categories with different values of $\kappa_p = -8/(3\sqrt{3})$, $16/(3\sqrt{3})$, and 0 for 32, 4, and 28 states, hence, giving the averaged nonzero value

$$\begin{aligned}\kappa &= \frac{1}{64}\left(-\frac{8}{3\sqrt{3}} \times 32 + \frac{16}{3\sqrt{3}} \times 4 + 0 \times 28\right) \\ &= -\frac{1}{\sqrt{3}} \approx -0.577\,.\end{aligned} \tag{S21}$$

The u3MC simulations of the full model $\mathcal{H}_{\text{BBQ-K}}$ in Eq. (1) of the main text revealed for the CSL a value of $|\kappa|$ which is reduced, but in the same order of magnitude.

### B. Correlations in momentum space

Correlations for magnetic moments in momentum space can be computed from the equal-time structure factor

$$S_\lambda(\mathbf{q}) = \left\langle \sum_{\alpha\beta} |m_\lambda{}_\beta^\alpha(\mathbf{q})|^2 \right\rangle, \tag{S22}$$

where $\langle \ldots \rangle$ represents an average over statistically-independent measurements, and the index $\lambda$ denotes the

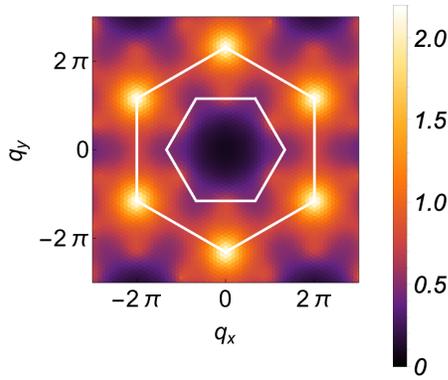

Figure S3. Quadrupolar structure factor, $S_Q(\mathbf{q})$, in the CSL phase, obtained by u3MC simulations at $T = 0.003$, corresponding to the middle panel of Fig. 4(d) of the main text.

channel for $\mathcal{A}$ matrices, dipoles, and quadrupoles $\lambda = \mathcal{A}$, S, and Q, respectively. Here, $m_\lambda{}^\alpha_\beta(\mathbf{q})$ is defined by

$$m_S{}^\alpha_\alpha(\mathbf{q}) = -i \sum_{\beta,\gamma} \epsilon^{\alpha\gamma}_\beta m_{\mathcal{A}}{}^\beta_\gamma(\mathbf{q}), \tag{S23}$$

$$m_Q{}^\alpha_\beta(\mathbf{q}) = -m_{\mathcal{A}}{}^\alpha_\beta(\mathbf{q}) - m_{\mathcal{A}}{}^\beta_\alpha(\mathbf{q}) + \frac{2}{3}\delta^\alpha_\beta \sum_\gamma m_{\mathcal{A}}{}^\gamma_\gamma(\mathbf{q}), \tag{S24}$$

where the Fourier transform of the $\mathcal{A}$ matrices is given as

$$m_{\mathcal{A}}{}^\alpha_\beta(\mathbf{q}) = \frac{1}{\sqrt{N_s}} \sum_i^{N_s} e^{i\mathbf{q}\cdot\mathbf{r}_i} \mathcal{A}^\alpha_{i\beta}. \tag{S25}$$

Our finite-temperature u3MC simulations revealed the existence of a CSL, showing diffuse scattering in the spin structure factor $S_S(\mathbf{q})$ for dipole moments, as depicted in Fig. 4(d) of the main text. As shown in Fig. S3, we confirmed that not only the dipolar, but also the quadrupolar structure factor $S_Q(\mathbf{q})$ is diffuse, indicating the absence of magnetic dipole and quadrupolar order in the CSL. Scattering features look very similar to the dipole ones, $S_S(\mathbf{q})$, since spin moments contain mainly dipole character which are implicitly captured by quadrupole components.

### C. Energy difference to the ground state

Figure 2(c) of the main text shows the normalized energy difference $\Delta E/N_s$ from the noncoplanar (NC) ordered state to the macroscopically degenerate CSL state at $\theta/\pi = 0.06$. The values have been obtained after energy minimization from randomly generated CSL configurations on the lattice.

In Fig. S4(a) we show the energy difference for a region in the phase diagram from $0.28 < \phi/\pi < 0.33$ and $0 < \theta/\pi < 0.2$, indicating that the finite-temperature CSL is expected to appear in an extended region in parameter space. In Fig. S4(b) we further plot the minimum value of the energy difference $\Delta E_{\min}$ at each $\theta$ along the green line in Fig. S4(a). We find that the energy difference vanishes at exactly the Kitaev point ($\theta = 0$), indicating that the CSL forms a submanifold of the degenerate manifold of the semiclassical Kitaev spin liquid.

## III. SEMICLASSICAL ANALOG OF THE $S = 1$ KITAEV SPIN LIQUID

We benchmark our method for the pure Kitaev model in absence of $J_1$ and $J_2$ interactions, and show in Fig. S5 the dipolar structure factor, $S_S(\mathbf{q})$, for the semiclassical analog of the AFM ($\theta/\pi = 0$) and FM ($\theta/\pi = 1$) Kitaev spin liquid. The diffuse structure is reminiscent to results seen from semiclassical solutions for the $S = 1/2$ model [16], and compares well to the structure observed at $T = 0.01$ in Fig. 4(d) of the main text.

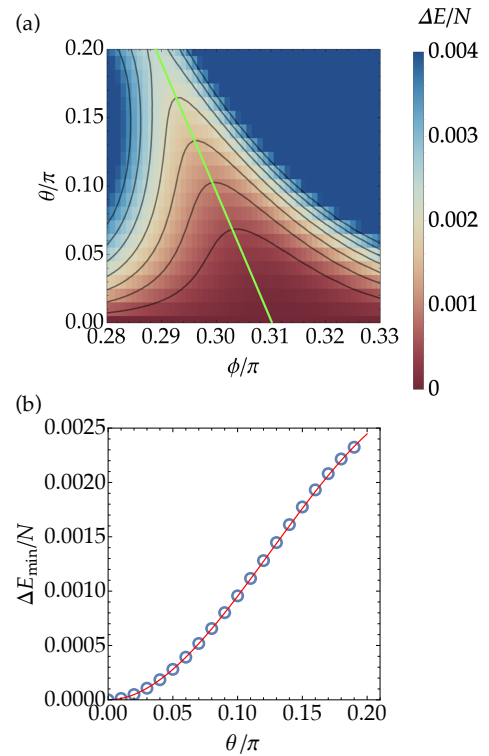

Figure S4. Normalized energy difference $\Delta E/N_s$ from the noncoplanar (NC) ordered state to the CSL. (a) Small values of $\Delta E/N_s$ between $0.28 < \phi/\pi < 0.33$ and $0 < \theta/\pi < 0.2$ imply that the finite-$T$ CSL exists in an extended region of the phase diagram. The green line connects the minimum values in the energy difference, $\Delta E_{\min}/N_s$, while the black ones are the energy contours. (b) $\Delta E_{\min}/N_s$ on the green line in (a) plotted as a function of $\theta$.

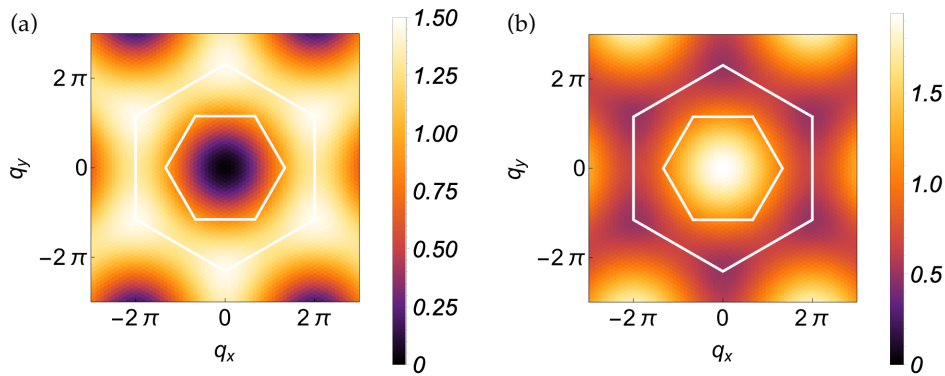

Figure S5. Dipole structure factor, $S_S(\mathbf{q})$, for the (a) AFM Kitaev spin liquid at $\theta/\pi = 0$ and (b) the FM Kitaev spin liquid at $\theta/\pi = 1$. Results were obtained from u3MC simulations for a $L = 24$ ($N = 1152$) sites cluster at $T = 0.01$.



\end{document}